\documentclass[journal]{IEEEtran} 

\usepackage{color}
\usepackage{balance}
\usepackage{graphicx}
\usepackage{enumerate}
\usepackage{subfigure}
\usepackage{algorithm}
\usepackage{algorithmic}
\usepackage{booktabs}
\usepackage{cite}
\usepackage{amssymb}
\usepackage{amsmath}
\usepackage{flushend}

\begin{document}

\title{Retroreflective Optical ISAC for 6G: Technologies, Applications and Future Directions}
	
\author{Tiantian Chu, Chen Chen, \IEEEmembership{Senior Member, IEEE}, Jia Ye, Xin Xiong, Sihua Shao, \IEEEmembership{Senior Member, IEEE}, \\
Zhihong Zeng, Dengke Wang, Fengli Yang, Guanjun Xu, and Harald Haas, \IEEEmembership{Fellow, IEEE}

\thanks{This work was supported in part by the National Natural Science Foundation of China under Grant 62501088 and Grant 62271091, in part by the National Science Foundation under Grant CNS-2431272, and in part by the Fundamental Research Funds for the Central Universities under Grant 2024CDJXY020. \textit{(Corresponding authors: Chen Chen; Sihua Shao)}}

\thanks{Tiantian Chu and Jia Ye are with the State Key Laboratory of Power Transmission Equipment Technology, School of Electrical Engineering, Chongqing University, Chongqing 400044, China (e-mail: 20241101050@stu.cqu.edu.cn; jia.ye@cqu.edu.cn).}

\thanks{Chen Chen, Zhihong Zeng, Dengke Wang, and Fengli Yang are with the School of Microelectronics and Communication Engineering, Chongqing University, Chongqing 400044, China (e-mail: c.chen@cqu.edu.cn; zhihong.zeng@cqu.edu.cn; dengke.wang@cqu.edu.cn; cqu\_yfl@cqu.edu.cn).}

\thanks{Xin Xiong and Sihua Shao are with the Department of Electrical Engineering, Colorado School of Mines, Golden, Colorado 80401, USA (e-mail: xin\_xiong@mines.edu; sihua.shao@mines.edu).}

\thanks{Guanjun Xu is with the Space Information Research Institute and Zhejiang Key Laboratory of Space Information Sensing and Transmission, Hangzhou Dianzi University, Hangzhou 310018, China (e-mail: gjxu@hdu.edu.cn).}

\thanks{Harald Haas is with the Department of Engineering, Electrical Engineering Division, Cambridge University, CB3 0FA Cambridge, UK (e-mail: huh21@cam.ac.uk).}
}

\markboth{}
{}

\maketitle

\begin{abstract}

Integrated sensing and communication (ISAC) has emerged as a key technological paradigm for sixth generation (6G) mobile networks, aiming to unify sensing and communication in a spectrally efficient and hardware lightweight manner. Radio frequency ISAC (RF-ISAC) is constrained by spectrum crowding, limited sensing resolution, and susceptibility to electromagnetic interference. In contrast, optical ISAC (O-ISAC) leverages the large bandwidth and short wavelength of optical carriers and is regarded as an important complement to RF-ISAC. However, conventional O-ISAC relies on natural optical reflections from target surfaces, which generate weak echoes that are highly dependent on surface materials, thereby limiting the achievable sensing range and sensing accuracy. This article introduces retroreflective optical ISAC (RO-ISAC), which alleviates these limitations by equipping targets with compact retroreflective modules. These modules return incident light approximately back to the source over a useful range of incidence angles, forming a well controlled double pass optical path with strong and stable echoes, and thereby further unlocking the application potential of O-ISAC. The conceptual architecture of RO-ISAC is presented together with the underlying retroreflection mechanism. Key enabling technologies for RO-ISAC systems are discussed, including channel modeling, waveform design, bidirectional transmission, and multi-target sensing and communication, with representative details and experimental validation. The suitability of RO-ISAC is analyzed in indoor, aerial, underwater, and satellite scenarios, and challenges and research directions related to mobility, cooperative networking, intelligent operation, and sustainable deployment are outlined, pointing toward robust and scalable RO-ISAC deployment in future 6G networks.

\end{abstract}


\section{Introduction}
Sixth generation (6G) wireless networks are expected to support massive connectivity, high data rates, and centimeter level localization, thereby enabling a wide range of intelligent applications such as industrial automation, extended reality, and large-scale environmental monitoring. However, the isolated evolution of communication and sensing has led to inefficient use of spectrum and hardware resources, and this drawback has become increasingly prominent in recent years. Integrated sensing and communication (ISAC) has been identified in the IMT-2030 framework as one of the core usage scenarios for 6G \cite{IMT2030ITU}. ISAC uses the same wireless signals both to deliver information to communication users and to sense surrounding targets and environments by processing the received echoes \cite{liu2022integrated}. By reusing a general waveform, hardware platform, and spectrum resource, ISAC enables hardware reuse, spectrum sharing, and functional reciprocity between communication and sensing.

Radio frequency ISAC (RF-ISAC) has made substantial progress in recent years, with many system architectures and prototypes proposed. It remains the mainstream solution for wide area coverage, yet increasingly severe spectrum congestion and the demand for higher sensing resolution have stimulated strong interest in optical approaches. Benefiting from available bandwidth that exceeds RF by several orders of magnitude and from the excellent collimation of laser diode (LD) beams, optical ISAC (O-ISAC) links can achieve data rates on the order of gigabits per second. The large bandwidth and high beam directivity also bring significant sensing gains \cite{liang2024integrated}. Under reasonable aperture and sampling conditions, O-ISAC can provide centimeter-level range resolution and microradian-level angular resolution \cite{wen2024optical}. In addition, by leveraging line-of-sight (LoS) propagation and flexible control of the beam direction, O-ISAC systems can effectively suppress interference from unintended light sources. These aspects highlight the significant application potential of O-ISAC.

\begin{figure*}[t!]
\centering
{\includegraphics[width=2\columnwidth]{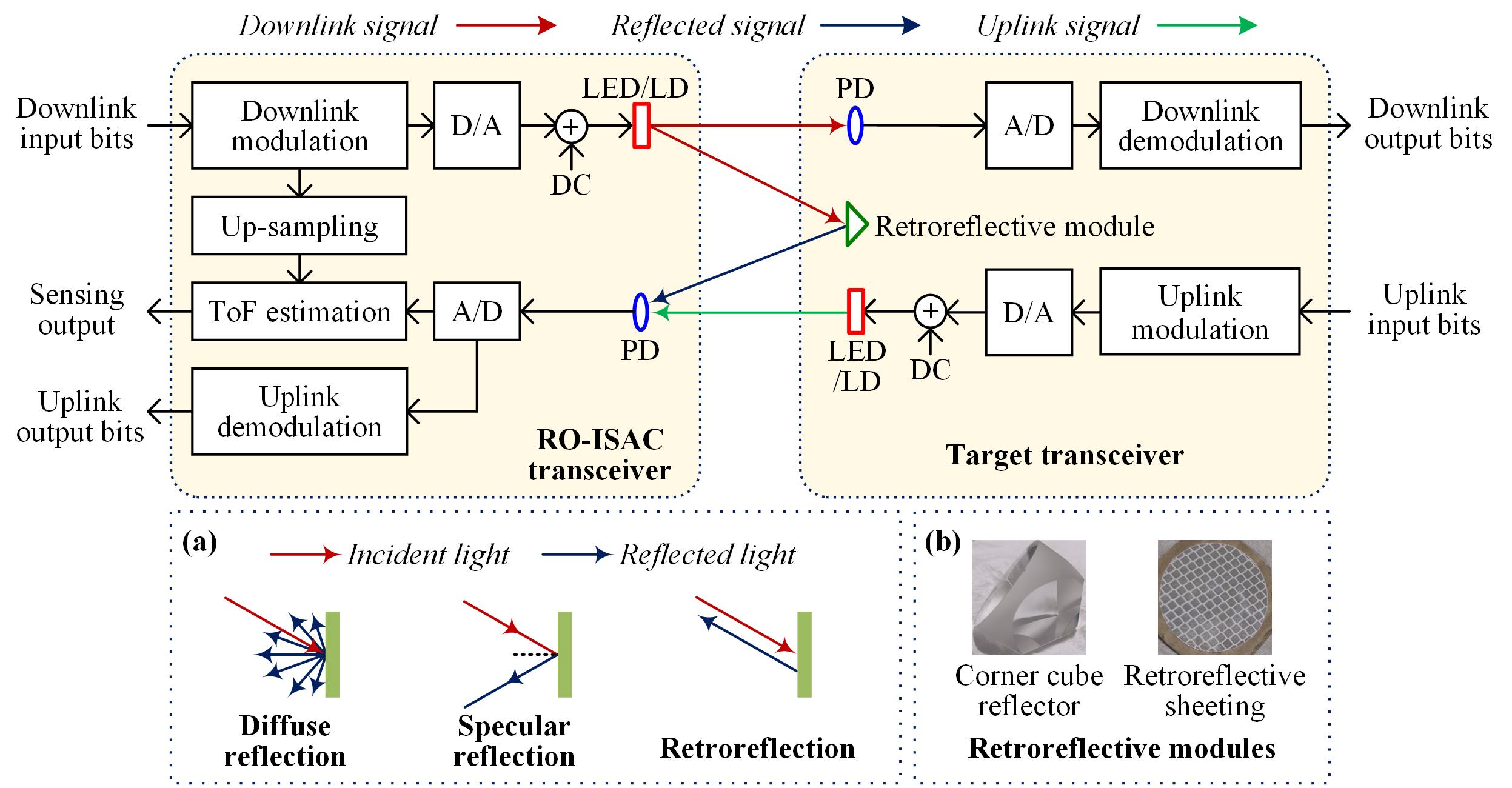}} 
\caption{Overall system Architecture of RO-ISAC. Insets: (a) illustration of three types of light reflection and (b) photos of two retroreflective modules.}
\label{fig:1Architecture}
\end{figure*}

Despite these advantages, conventional O-ISAC schemes still rely on natural reflections from surfaces such as walls, ceilings, and various objects. In such environments, the strength and stability of the echo signal are largely constrained by the surface material and geometry, which often leads to weak returns and strong sensitivity to color and roughness. It should be emphasized that spatial resolution is not the only factor determining sensing performance. Echo quality and noise level jointly determine the achievable signal-to-noise ratio, which directly affects ranging and localization accuracy. Under these limitations, achieving stable and reliable sensing over large detection distances is highly challenging. More controllable reflection architectures are required to fully unlock the potential advantages of O-ISAC.

Motivated by these limitations, this article introduces retroreflective optical ISAC (RO-ISAC), which equips targets with compact retroreflective modules. These modules return incident light approximately toward its source over a useful range of incidence angles, so the dominant sensing and communication path becomes a well controlled double pass link governed by the retroreflector rather than by uncontrolled ambient surfaces. The engineered retroreflection yields a much stronger and more predictable echo, supporting reliable sensing over longer distances. Combined with the large bandwidth, short wavelength, and narrow beams of optical carriers, RO-ISAC can achieve higher data rates, finer sensing resolution, and tighter spatial confinement than RF-ISAC, while providing stronger and more controllable returns than conventional O-ISAC. We investigates the system architecture, enabling technologies, and representative application scenarios of RO-ISAC, and outlines key challenges and research directions toward its practical deployment in future 6G wireless networks. To the best of our knowledge, it is the first overview article devoted specifically to RO-ISAC.

\section{System Architecture of RO-ISAC}
The overall system architecture of RO-ISAC is illustrated in Fig.~\ref{fig:1Architecture}. The RO-ISAC transceiver integrates downlink transmission, uplink reception, and sensing processing within a single unified optical front end, while the target transceiver integrates downlink reception, a retroreflective module, and uplink transmission within a compact terminal.

\textbf{RO-ISAC Transceiver:} The downlink input bits are first modulated into a real-valued ISAC signal. This digital signal is converted into an analog waveform through digital-to-analog (D/A) conversion and combined with a direct-current (DC) bias to form the analog driving signal to be transmitted, which drives a light-emitting diode (LED) or LD to generate the optical signal. The optical signal propagates to the target and illuminates the retroreflective device mounted on it. The photodiode (PD) simultaneously receives the downlink signal reflected from the target and the uplink signal transmitted by the target side. The received analog electrical signal is converted into the digital domain through analog-to-digital (A/D) conversion and is then used for two tasks. One task is uplink demodulation to recover the uplink output bits, and the other is correlation with the upsampled transmit signal to extract time-of-flight (ToF) and further yield the sensing output.

\textbf{Target Transceiver:} The target uses its PD to detect the incident downlink light and convert it into an electrical signal, from which the downlink bits are recovered after A/D conversion and downlink demodulation. The target also integrates a retroreflective module, such as a corner-cube reflector (CCR) or retroreflective sheeting, which redirects the received downlink light approximately along the original path back to the RO-ISAC transceiver and thereby provides a stable echo for distance measurement and target sensing at the RO-ISAC transceiver. When uplink communication is required, the target maps the uplink input bits to modulation symbols and generates the uplink driving signal through D/A conversion and DC biasing.

The point-to-point RO-ISAC architecture can be used to measure the distance between the transceiver and the target. Building on this capability, the system can be extended to support more advanced sensing tasks, including two- and three-dimensional localization as well as velocity estimation. In localization scenarios, multiple RO-ISAC transceivers are typically deployed, each estimating its own distance to the target. These distance measurements are then jointly processed to obtain the target position. For velocity estimation, the system tracks changes in the target position over a given time window and computes the speed from the corresponding displacement and time interval.

\begin{figure}[!t]
\centering
{\includegraphics[width=1\columnwidth]{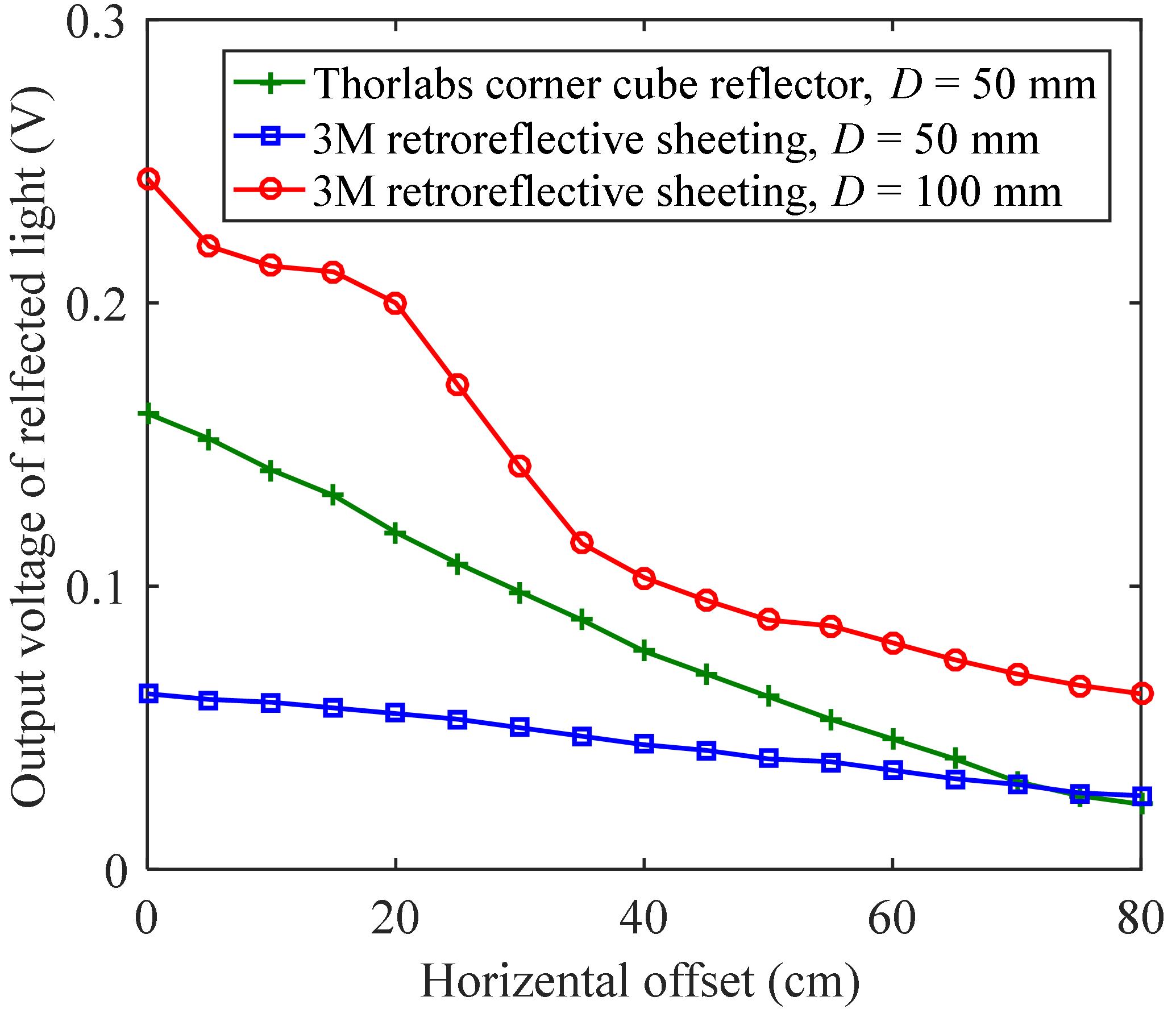}} 
\caption{Retroreflection performance of Thorlabs CCR and 3M retroreflective sheeting.}
\label{fig:2Retroreflection}
\end{figure}

The benefits of RO-ISAC are closely linked to the reflection mechanism. As illustrated in Fig.~\ref{fig:1Architecture}(a), diffuse reflection spreads incident light over many directions, which causes severe path loss and produces echoes that depend strongly on surface color and roughness. Specular reflection follows the rule that the angle of reflection equals the angle of incidence, so the main reflected lobe leaves along the specular direction rather than returning toward the source. When the transmitter and receiver are at the same position or placed close together, most of this energy bypasses the receiver unless the surface orientation falls within a narrow angular range. If the transmitter and receiver are separated to capture the specular direction, the overall size of the terminal increases. Retroreflection avoids these limitations by sending the incident beam back approximately along the incoming path over a range of incidence angles, which yields a stronger and more stable echo for a given transmit power. These properties can be realized with practical retroreflective modules. Fig.~\ref{fig:1Architecture}(b) shows two representative examples, the CCR and the 3M retroreflective sheeting. The measurement results in Fig.~\ref{fig:2Retroreflection} show that both types maintain a high reflected level as the horizontal offset increases, and that enlarging the aperture of the retroreflective sheeting further enhances the return power and its robustness to misalignment. Retroreflective modules provide a simple and effective way to realize strong and geometry-robust optical echoes for RO-ISAC systems.

\section{Enabling Technologies for RO-ISAC}
This section discusses several key technologies that transform the basic RO-ISAC concept into a practical system, including channel modeling, waveform design, bidirectional transmission, and mechanisms for multi-target communication and sensing.

\subsection{Channel Modeling}

\begin{figure*}[htbp]
\centering
{\includegraphics[width=2\columnwidth]{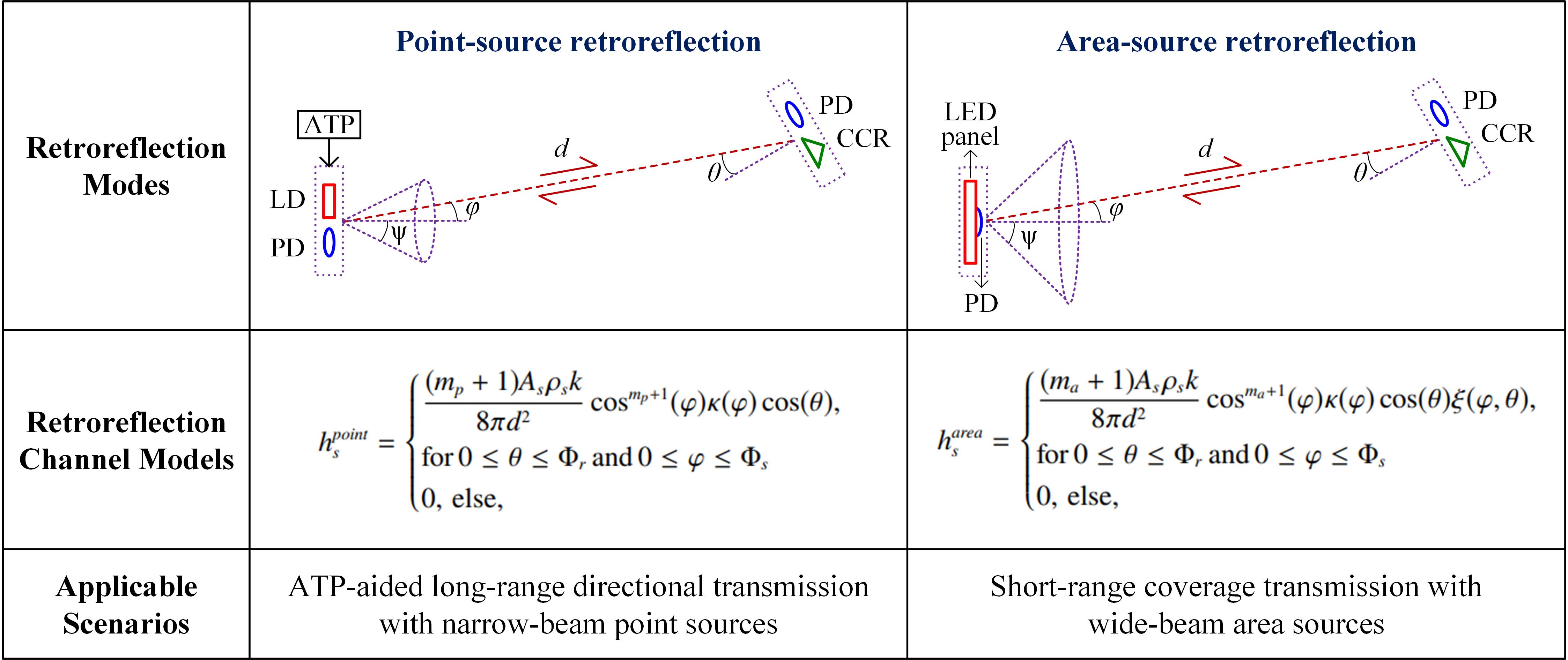}} 
\caption{Point-source and area-source RO-ISAC retroreflection channel models. ${m}_{p}$ and ${m}_{a}$: the corresponding Lambertian emission order; $d$: distance between the RO-ISAC sensing transceiver and the CCR; $\phi$: irradiance angle of the emitted beam with respect to the transceiver boresight; $\theta$: incidence angle at the CCR; $A_{s}$: active area of the sensing PD; $\rho_{s}$: responsivity of the sensing PD; $\Phi_{s}$: half-angle FOV of the sensing PD; $k$: reflectance of the CCR; $\Phi_{r}$: half-angle FOV of the CCR; $\kappa(\phi)$: geometric factor describing the impact of the incidence angle on the received reflected signal; $\xi(\phi,\theta)$: effective reflecting ratio of the CCR with respect to the area source.}
\label{fig:3Model}
\end{figure*}

Channel modeling for RO-ISAC mainly focuses on the retroreflective sensing path. In the considered architecture, the downlink optical signal is emitted by the RO-ISAC transceiver, illuminates the retroreflective device located at the target, and then returns to the RO-ISAC transceiver along approximately the same geometric path. The effective channel gain of this double pass link depends on multiple factors. A tractable channel model is therefore essential for link budget evaluation and parameter optimization. Narrow-beam LDs and wide-beam LEDs are two commonly used types of light sources, and RO-ISAC sensing channel models for the corresponding point-source and area-source configurations have been proposed, as illustrated in Fig.~\ref{fig:3Model} \cite{cui2024retroreflective}.

\textbf{Point-Source Retroreflection Model:} In the point-source configuration, a narrow-beam LD is steered toward the CCR with the aid of acquisition, tracking, and pointing (ATP). The transmitted power is concentrated around the beam axis, so the echo received at the sensing PD is mainly determined by the round trip propagation distance, the aperture of the CCR, and the overlap between the beam and the fields of view (FOVs) of the CCR and the PD. The point source channel model captures these effects through distance dependent attenuation, Lambertian emission characteristics, and angle weighting terms. Because the sensing signal must be retroreflected back to the PD at the RO-ISAC transceiver, an additional cosine factor appears in the channel gain and the effective propagation distance is doubled. When the beam is well aligned, this mode provides a strong retroreflected signal and supports long-range, high signal-to-noise ratio sensing, whereas misalignment causes the gain to drop rapidly, highlighting the critical role of accurate ATP in this regime.

\textbf{Area-Source Retroreflection Model:} In the area-source configuration, a wide-beam emitter such as an LED panel illuminates the scene, and only the portion of its radiation that falls within the CCR FOV contributes to an effective retroreflected signal. The returning light is further limited by the aperture and FOV of the sensing PD. Consequently, in addition to the distance and angle dependent factors that also appear in the point-source case, the area-source channel model introduces an effective reflecting ratio that quantifies the fraction of the illuminated panel area that actually participates in retroreflection. Compared with the narrow-beam mode, the area-source mode exhibits a lower peak gain but is more tolerant to misalignment and orientation changes. It is therefore well suited to short-range sensing and wide-coverage indoor deployments, particularly in scenarios with a large number of retroreflective tags.

Retroreflection oriented channel models characterize the channel gain properties of RO-ISAC links in a concise and effective manner. These models support the theoretical analysis and optimization of RO-ISAC systems at the system level.

\subsection{Waveform Design}

The communication and sensing performance of RO-ISAC systems depends strongly on the underlying integrated waveform, so the design of high-performance waveforms is of central importance. In practical deployments, different scenarios place different emphasis on data rate, sensing accuracy, and implementation complexity, which naturally leads to three representative classes of RO-ISAC waveforms. These include sensing-centric designs that focus on improving ranging and localization capability, communication-centric designs that prioritize throughput and robustness, and hybrid designs that strike a balance between the two. 

\textbf{Sensing-Centric Waveform Designs:}
Sequences with good autocorrelation properties, such as pseudo random sequences, are transmitted as pulse trains or pulse position modulated signals \cite{wen2023pulse}. With the strong echo of the retroreflective link, these signals can use low duty cycles and high peak power to obtain narrow main lobes and low sidelobes, which benefits delay estimation, multi target resolution, and multipath suppression. The cost is limited symbol resources and spectral efficiency for data transmission.

\textbf{Communication-Centric Waveform Designs:}
This line of design adopts conventional optical communication schemes, with orthogonal frequency-division multiplexing (OFDM) as a representative choice \cite{wen2025free}. High-order constellations and multi-carrier modulation enable high data rates over the retroreflective link, but the resulting waveform has weak correlation structure in time and a relatively high peak to average power ratio (PAPR), which stresses the linear region of LEDs or LDs. Suitable clipping in the time domain can reduce peaks while maintaining good communication performance and acceptable sensing quality.

\begin{figure}[!t]
\centering
{\includegraphics[width=1\columnwidth]{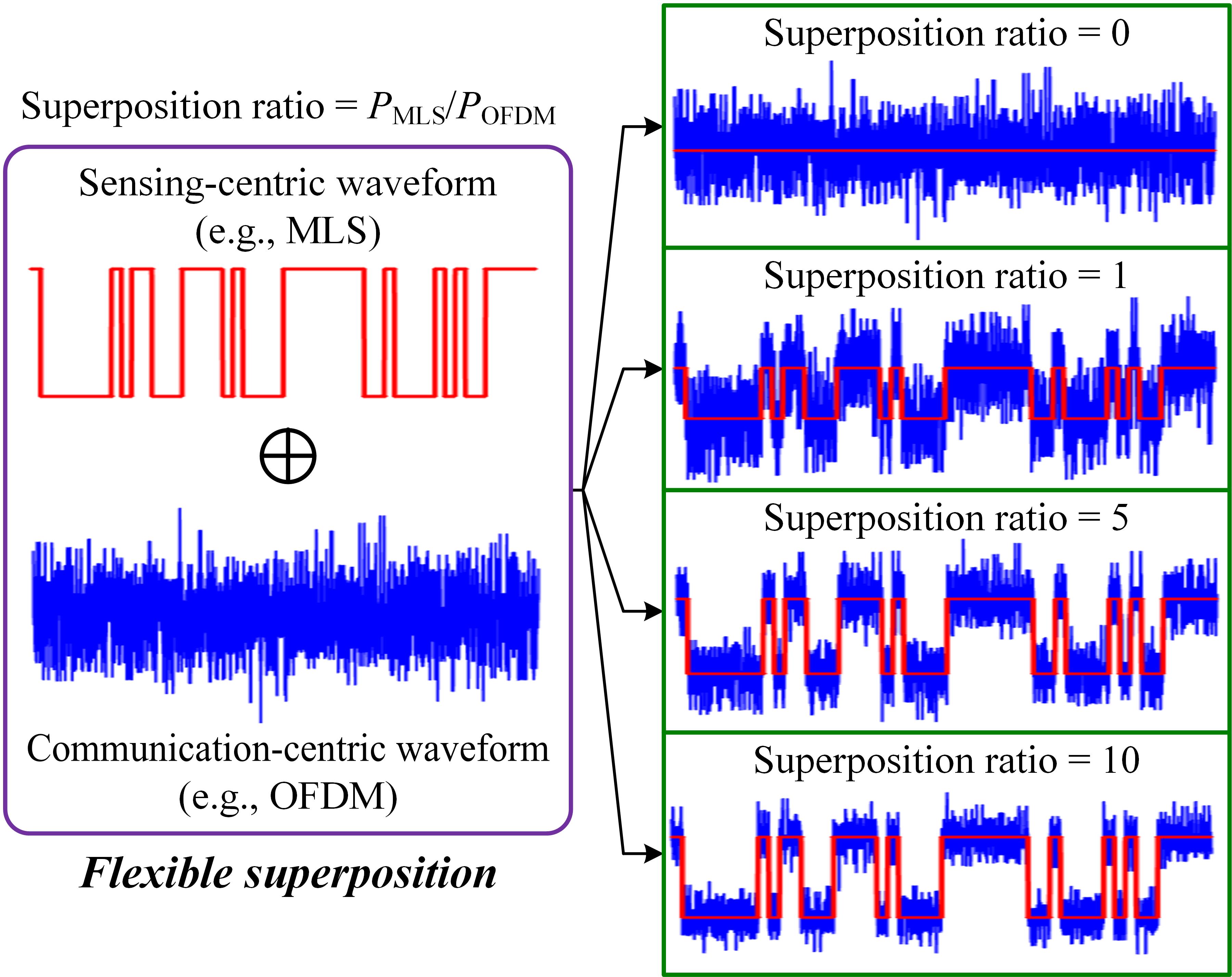}} 
\caption{Illustration of the balanced communication–sensing waveform designs with different superposition ratios.}
\label{fig:4Waveform}
\end{figure}

\begin{figure}[!t]
\centering
{\includegraphics[width=1\columnwidth]{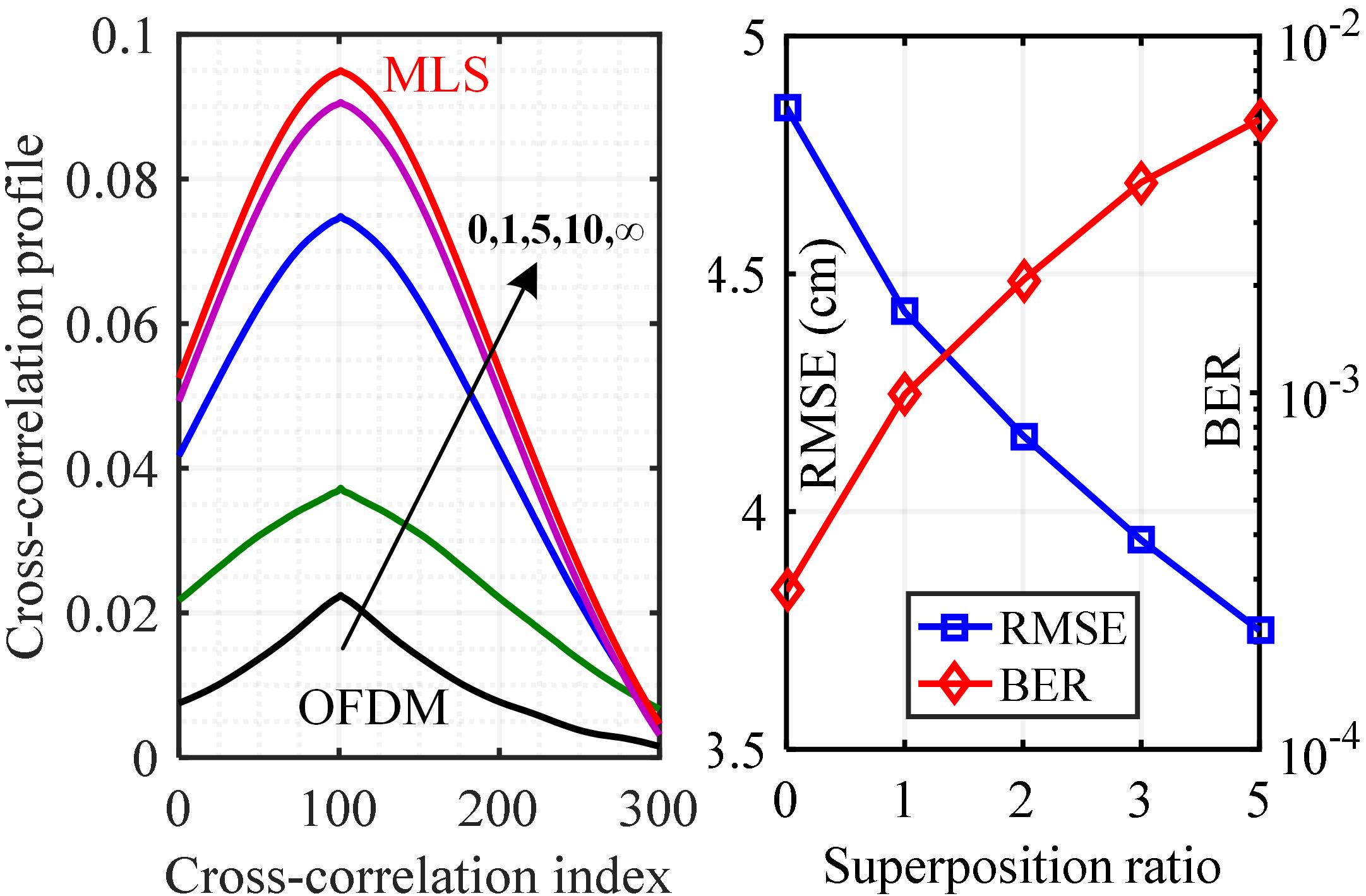}} 
\caption{Cross-correlation, RMSE, and BER performance of the OFDM-MLS hybrid waveform with different superposition ratios.}
\label{fig:5Pperformance}
\end{figure}

\textbf{Communication–Sensing Trade-Off Waveform Designs:}
A balance between sensing-centric and communication-centric designs can be achieved by embedding structured pulse sequences into a high-rate carrier \cite{Du2025Flexible}. In the OFDM waveform with embedded maximum length sequence (MLS) shown in Fig.~\ref{fig:4Waveform}, a bipolar zero-mean OFDM signal is superimposed with a bipolar MLS sequence in the power domain to form an integrated waveform. As indicated in Fig.~\ref{fig:5Pperformance}, adjusting the superposition ratio between the MLS and OFDM components realizes a flexible trade-off between communication and sensing performance, although the MLS part itself does not convey additional user data.

\subsection{Bidirectional Transmission}

Practical RO-ISAC systems need bidirectional links to exchange information between the transceiver and the target. Unlike conventional optical wireless systems, the PD at the RO-ISAC transceiver receives a superposition of the uplink signal from the target and the retroreflected downlink signal, which creates self interference. Reliable operation therefore requires mechanisms to separate or suppress these components, for which time division duplexing (TDD) \cite{wang2025bidirectional} and wavelength division duplexing (WDD) \cite{chen2025full} are two practical options.

\textbf{Time Division Duplexing:} Downlink and uplink are assigned to different time slots in a frame. In the downlink slot, the transceiver transmits a joint communication and sensing waveform and uses the retroreflected signal for target sensing. In the uplink slot, the target modulates its data and the transceiver processes the received signal only as uplink. A guard interval between the two slots allows the strongest downlink echoes to decay before uplink reception, and its length can be set according to the estimated maximum range to avoid unnecessary overhead for nearby targets. TDD uses a single optical wavelength and simple hardware, but operates in a half-duplex manner, so the effective data rate and sensing update rate are constrained by the frame structure.

\textbf{Wavelength Division Duplexing:} WDD separates downlink and uplink in the wavelength domain. The downlink may use a blue source and the uplink a green source, while optical filters at the transceiver and the target split the received light into two bands. The sensing function relies on the echo of the downlink wavelength, whereas the other wavelength carries uplink data. In this way, downlink, uplink, and sensing can proceed simultaneously with very weak self interference. WDD supports true full-duplex transmission and higher aggregate data rates, at the cost of additional light sources, wavelength selective filters, and more complex alignment and calibration of the optical front ends.

\subsection{Multi-Target Sensing and Communication}

In many RO-ISAC deployments, a single transceiver must simultaneously track multiple targets and support communication with multiple terminals rather than interacts with an isolated retroreflective node. Typical examples include indoor environments with dense asset tags, roads with many vehicles carrying retroreflective plates, and factory floors where robots and workers are both equipped with tags. All targets are illuminated by the same RO-ISAC waveform and each CCR or retroreflective sheet returns a copy of the incident signal. This coupling makes multi-target sensing and multi-target communication tightly related design problems.
    
\textbf{Multi-Target Sensing:} Multi-target sensing exploits both spatial multiplexing and echo processing. In the optical front end, different illumination directions, receiver FOVs, or detector apertures can be configured so that the coverage region is partitioned into several spatial sectors and each target mainly contributes to the retroreflected signal within a subset of these sectors. This spatial multiplexing reduces mutual coupling among targets and enables parallel estimation of range. In the time domain, the strong retroreflected link yields delay profiles with well separated peaks, where nearby strong targets may mask weaker ones. By successively estimating the dominant echoes and subtracting their reconstructed contributions, the residual signal reveals weaker or more distant targets. Combining spatial partitioning with such echo serial interference cancellation significantly enhances multi-target sensing performance over a wide-range of target distances and reflection strengths.

\textbf{Multi-Target Communication:}
With a wide illumination beam and a general integrated waveform, the RO-ISAC transceiver can serve multiple user terminals simultaneously over the same optical link. Orthogonal multiple access (OMA) can separate terminals in time or frequency, for example by assigning disjoint OFDM subcarrier groups or time slots within one RO-ISAC frame so that each terminal only decodes the data mapped to its own resources. Non-orthogonal multiple access (NOMA) can further improve spectral efficiency by allowing controlled overlap in time–frequency resources and separating users through power-domain or code-domain processing at the receiver. These multiple-access schemes can be embedded into the integrated RO-ISAC waveform so that communication symbols share a unified time–frequency grid with sensing probes. In all cases, the multiple-access design should be coordinated with the sensing function to ensure that resource allocation maintains sufficient probing energy and update rate for every illuminated target.

\section{Application Scenarios of RO-ISAC}

\begin{figure*}[htbp]
\centering
{\includegraphics[width=2\columnwidth]{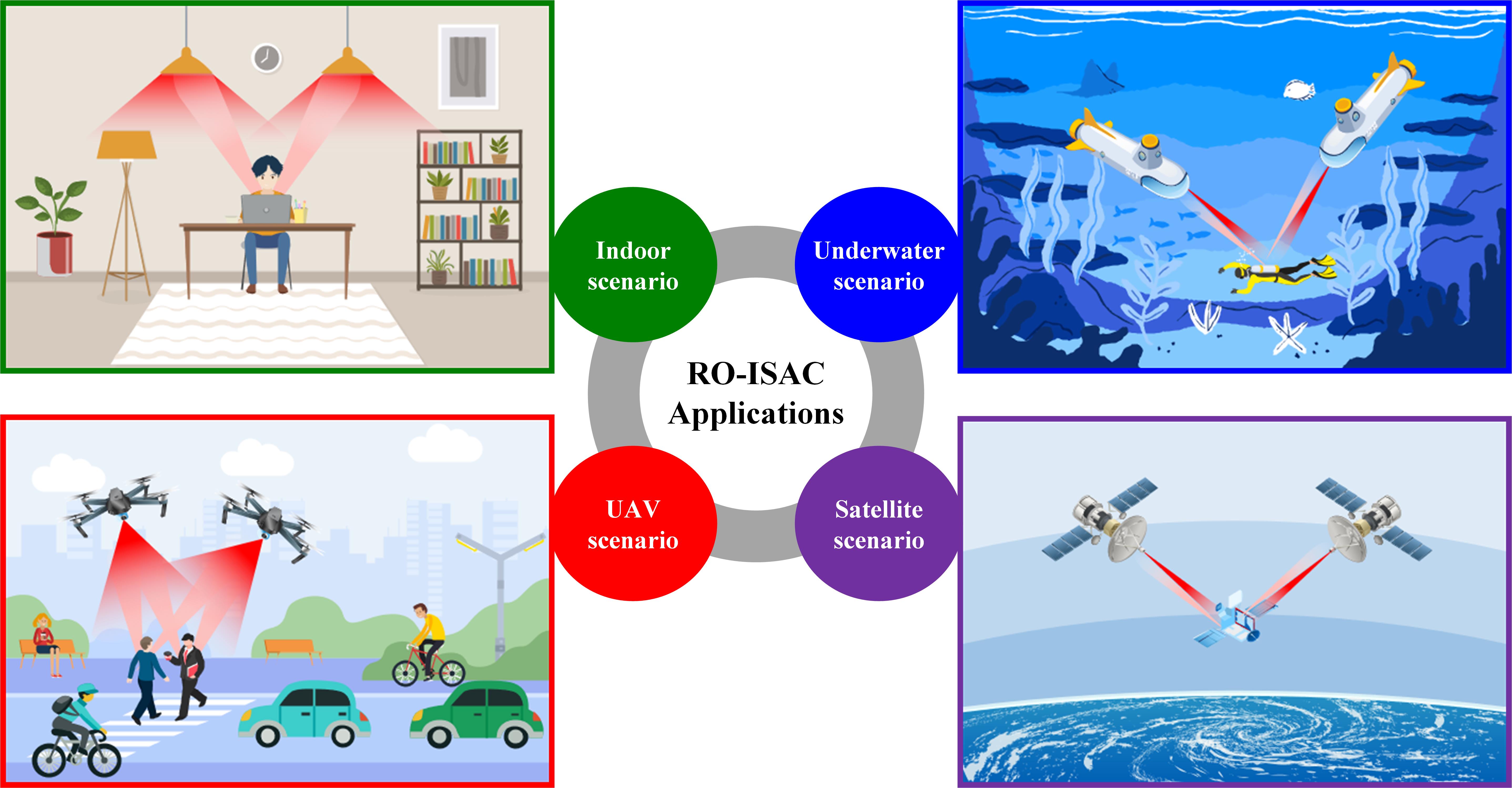}} 
\caption{Illustration of RO-ISAC application scenarios.}
\label{fig:6Applications}
\end{figure*}

RO-ISAC is not tied to a single deployment style. By combining strong retroreflection with joint communication and sensing, it can be tailored to very different environments, from compact indoor spaces to unmanned aerial vehicle (UAV), underwater, and satellite links. This section outlines several representative application domains and highlights how the properties of RO-ISAC map naturally onto their requirements.

\subsection{Indoor Applications}

Indoor environments such as factories, warehouses, shopping malls, and hospitals are natural candidates for O-ISAC \cite{shi2024experimental}. Ceiling-mounted LED panels or LD–PD units can act as RO-ISAC transceivers, while assets, robots, or people carry small retroreflective tags. In this setting the system typically operates in an area-source mode, where wide-beam illumination from the ceiling provides coverage and CCR or retroreflective sheeting on the tags send back a concentrated echo that can be used for both communication and sensing. Compared with traditional optical ISAC that relies on diffuse reflections from walls and objects, RO-ISAC provides a much stronger and more stable return from each tagged entity. This enables accurate ranging and localization over longer distances within the room, even when the environment is cluttered or the background surfaces are poorly reflective. At the same time, the narrow return beam improves privacy and interference isolation. Communication and sensing are largely confined to the link between a transceiver and its tag, which is attractive for applications such as patient tracking in hospitals, inventory management in warehouses, or human–robot collaboration on factory floors.

\subsection{UAV Applications}

UAV scenarios feature high mobility, rapid topology changes, and strict payload constraints \cite{meng2024uav}, which creates conditions where RO-ISAC is particularly attractive. A UAV mounted RO-ISAC transceiver can interrogate retroreflective markers placed on the ground, on building facades, or on other moving platforms. In a typical point-source configuration, the UAV uses a narrow-beam LD with ATP to illuminate the target, while the CCR on the target returns a strong echo that carries the sensing information. The retroreflected beam provides a stable and high resolution ranging signal that supports tasks such as precision landing, infrastructure inspection, and cooperative flight. Retroreflective targets can remain very simple and lightweight because no active light source is required, which is well suited to disposable beacons or low-cost auxiliary sensors. The strong spatial confinement of the optical link also reduces interference to and from other wireless systems and eases spectrum coexistence in crowded airspace. 

\subsection{Underwater Applications}

In oceanographic and environmental monitoring, high-precision localization and sensing enabled by ISAC are crucial for underwater Internet of Things (IoT) networks \cite{jehangir2024isac}. Underwater environments suffer from severe RF attenuation and complex acoustic propagation, which makes optical links an attractive complement for short-to-medium-range connectivity. RO-ISAC can be employed between autonomous underwater vehicles (AUVs) and retroreflective stations deployed on the seabed, on buoys, or on infrastructure such as pipelines and offshore platforms. Blue and green wavelengths are typically preferred in order to exploit the optical transmission window in water. A retroreflective station returns a strong and directionally confined echo that is less affected by multipath scattering than conventional diffuse reflection links. The RO-ISAC transceiver on the AUV can obtain precise range measurements to support navigation and docking, while simultaneously exchanging status information or sensor data with the station. Because the retroreflective unit is passive or semi-passive, it can be sealed for long-term deployment without frequent maintenance or large power supplies, which is attractive for deep-sea monitoring and long-duration environmental sensing. Key design considerations include coping with turbidity and biofouling, which reduce retroreflective efficiency, and selecting waveforms that remain robust under strong scattering and background light from the surface.

\subsection{Satellite Applications}

With the rapid deployment of large constellations of small satellites and CubeSats in low Earth orbit, compact and high-throughput inter-satellite links based on optical technology are becoming increasingly important for data relay and coordination in the space segment \cite{safi2025cubesat}. At the upper end of the range and mobility spectrum, RO-ISAC can support satellite-to-ground and satellite-to-airborne links. The satellite acts as a highly directional RO-ISAC transceiver equipped with narrow-beam optical terminals and precise pointing mechanisms, while ground stations, airborne platforms, or large moving objects carry retroreflective arrays. The link typically operates in a point-source mode over very long distances and benefits from the strong focusing and self-alignment properties of retroreflection. The combination of an optical carrier and a retroreflective return enables extremely high data rates together with centimeter level ranging accuracy, which is attractive for precise orbit determination, cooperative navigation, and high-throughput feeder links in future 6G satellite networks.

\section{Challenges and Future Directions}

RO-ISAC introduces a new paradigm in which downlink and uplink transmission are tightly integrated with sensing on a retroreflective link, but significant challenges remain on the road to practical deployment. Issues related to mobility, coordination among many nodes, algorithmic intelligence, and sustainable terminal operation still require careful study and system level validation. This section summarizes key open problems and discusses promising directions for advancing RO-ISAC toward robust and scalable use in future optical wireless networks.

\subsection{Mobile RO-ISAC}
Mobility is both a key driver and a major challenge for RO-ISAC. In indoor environments, users carrying retroreflective tags may walk, rotate, or be intermittently blocked, while in outdoor and aerial scenarios, UAVs and vehicles introduce fast and often unpredictable motion. Although retroreflection provides a certain degree of self alignment, the effective channel gain remains highly sensitive to relative position, orientation, and blockage. When the link operates in a narrow-beam point-source mode, even small pointing errors or rapid attitude changes can cause deep fades or complete link outages.

Future mobile RO-ISAC systems will therefore require tighter integration between optical transceivers and platform motion control. This includes fast acquisition and tracking loops, prediction of motion based on inertial sensors or historical trajectories, and waveform designs that remain robust under time-varying Doppler condition. Another open issue is how to maintain joint communication and sensing performance under mobility. For example, a waveform optimized for high data rate may not provide sufficient sensing resolution when the channel varies within a frame, whereas a sensing-oriented probing pattern may incur excessive overhead at high speeds. The development of mobility-aware frame structures, adaptive duplexing schemes, and cross-layer controllers that jointly account for motion, link quality, and sensing accuracy is an important direction for future work.

\subsection{Cooperative RO-ISAC}
Most current RO-ISAC prototypes focus on a single transceiver and one or several retroreflective targets. However, many envisioned 6G scenarios such as smart factories, intelligent transportation systems, and large-scale indoor positioning will require networks of RO-ISAC nodes that operate in a coordinated manner. Cooperative RO-ISAC extends the concept from a point-to-point link to a distributed sensing and communication fabric, where multiple transceivers share information and jointly illuminate and observe retroreflective devices.

Such cooperative operation enables multi-static sensing with improved accuracy and coverage, handover between transceivers to support user mobility, and spatial reuse through coordinated beam scheduling. At the same time, it introduces several new challenges. Transceivers must exchange control information and possibly raw or processed sensing data over backhaul links, which leads to latency and synchronization constraints. Interference among overlapping beams must be managed, especially when multiple nodes interrogate nearby targets. A key topic for future research is the design of scalable protocols that coordinate which node transmits, which node receives, and how sensing information is fused, while avoiding excessive overhead on the control plane. Hybrid architectures that combine RO-ISAC with RF backhaul or with reconfigurable intelligent surfaces (RISs) may further enhance coverage and robustness, but they also increase the overall system complexity.

\subsection{Intelligent RO-ISAC}
The strong coupling between geometry, channel conditions, and sensing outputs in RO-ISAC makes the system an attractive candidate for intelligent optimization. Machine learning can be used to predict channel quality, select beams, tune waveform parameters, and interpret sensing data. For example, a transceiver can learn from historical measurements how the retroreflected power varies with user trajectories and then proactively adjust its beam steering and power allocation. Learning-based detectors can also distinguish between different types of targets or motion patterns based on joint communication and sensing observations.

However, moving from isolated learning modules to a truly intelligent RO-ISAC architecture is far from trivial. Training data are expensive to collect in the optical domain, particularly under diverse environmental conditions such as fog, rain, or underwater turbidity. Models trained in one scenario may generalize poorly to others, and purely data-driven approaches may struggle to capture rare but critical events such as abrupt blockage or hardware failure. Future research will likely focus on hybrid methods that combine physics-based models with data-driven refinement, transfer learning across scenarios, and online adaptation under limited feedback. Another promising direction is semantic level design, in which intelligent RO-ISAC nodes adapt their operation according to the importance or urgency of the sensed information and transmitted data, which is especially relevant for safety-critical and low-latency applications.

\subsection{Integration of RO-ISAC with Wireless Power Transfer}
In many RO-ISAC oriented scenarios, the retroreflective terminals are envisioned as small and inexpensive devices that operate for long periods with minimal human intervention. Typical examples include tags attached to industrial equipment, structural elements, underwater installations, or distributed environmental sensors. Although the retroreflective elements themselves are passive, the associated electronics such as sensors, controllers, and uplink modulators still require a stable energy supply. Relying solely on primary batteries leads to high maintenance costs and is often infeasible in remote, harsh, or densely deployed environments. This situation naturally connects RO-ISAC to the broader idea of integrating communication, sensing, and power transfer, and suggests using the optical downlink not only as an information and probing carrier but also as an energy source.

Integrating RO-ISAC with wireless power transfer aims to build an optical framework where a single retroreflected link can both sustain device operation and support joint communication and sensing \cite{Chu2025Revolutionizing}. On the transceiver side, the main task is to shape waveforms and beam patterns so that a suitable DC component is available for energy harvesting while the remaining alternating current (AC) component maintains adequate performance for data transmission and high-resolution sensing under eye safety and hardware constraints. On the terminal side, compact front ends that combine photovoltaic elements with CCRs or retroreflective sheeting might divide the received optical power between energy harvesting and retroreflection without causing a large loss in ranging accuracy or uplink reliability. At the system level, resource management and end-to-end energy efficiency models are needed to decide how power, communication, and sensing resources are allocated across many devices.

\section{Conclusions}

RO-ISAC is expected to enable joint high-rate communication and high-precision sensing over extended distances in future optical wireless networks with compact retroreflective tags. This article has introduced the fundamentals of RO-ISAC, including the retroreflection mechanism and system architecture, and has discussed key enabling technologies such as channel modeling, waveform design, bidirectional transmission, and multi-target sensing and communication. Representative indoor, aerial, underwater, and satellite scenarios have been outlined to illustrate how RO-ISAC can be embedded into practical 6G applications. Remaining challenges in mobile operation, cooperative networking, intelligent configuration, and integration with wireless power transfer define a rich research agenda. This article aims to stimulate interests and investigations on the future development of RO-ISAC.

\bibliographystyle{IEEEtran}
\bibliography{IEEEabrv,mylib}

\end{document}